\documentclass[
reprint,
superscriptaddress,
amsmath,
amssymb,
aps,
prl,
]{revtex4-2}
\usepackage{ragged2e}
\usepackage{caption}
\usepackage[caption=false]{subfig}
\usepackage{ragged2e}
\DeclareCaptionJustification{justified}{\justifying}
\usepackage[colorlinks,allcolors=cyan]{hyperref}
 
\usepackage{graphicx}
\usepackage{dcolumn}
\usepackage{bm}
\usepackage[capitalise]{cleveref}
\usepackage{physics}
\usepackage{xcolor}
\usepackage{comment}
\usepackage{bbold}
\usepackage{mathtools}
\usepackage{dsfont}
\usepackage{bbold}
\usepackage{verbatim}
\newif\ifSM
\SMtrue
\newif\ifmainText
\mainTexttrue 

\begin{document}
\def\lc{\left\lfloor}   
\def\rc{\right\rfloor}
\setlength{\intextsep}{10pt plus 2pt minus 2pt}

\ifmainText
\title{Emergent Berezinskii-Kosterlitz-Thouless deconfinement in super-Coulombic plasmas}
 
\author{Ayush De}
 \affiliation{Racah Institute of Physics, The Hebrew University of Jerusalem, Jerusalem 91904, Israel}
 \author{Leo Radzihovsky}
 \affiliation{Department of Physics, University of Colorado, Boulder, Colorado 80309, USA}
 \author{Snir Gazit}
 \affiliation{Racah Institute of Physics, The Hebrew University of Jerusalem, Jerusalem 91904, Israel}
 \affiliation{The Fritz Haber Research Center for Molecular Dynamics, The Hebrew University of Jerusalem, Jerusalem 91904, Israel}
 
\date{\today}

\begin{abstract}
We study the statistical mechanics of two-dimensional ``super-Coulombic" plasmas, namely, neutral plasmas with power-law interactions longer-ranged than Coulomb. To that end, we employ numerically exact large-scale Monte Carlo simulations. Contrary to naive energy-entropy arguments, we observe a charge confinement-deconfinement transition as a function of temperature. Remarkably, the transition lies in the Berezinskii-Kosterlitz-Thouless (BKT) universality class. Our results corroborate recent dielectric medium and renormalization group calculations predicting effective long-scale Coulomb interactions in microscopically super-Coulombic gases. We explicitly showcase this novel dielectric screening phenomenon, capturing the emergent Coulomb potential and the associated crossover length scale. This is achieved by utilizing a new test charge based methodology for determining effective inter-particle interactions. Lastly, we show that this Coulomb emergence and the associated BKT transition occur universally across generic interactions and densities.
\end{abstract}
\maketitle

{\it Introduction --} Berezinskii-Kosterlitz-Thouless (BKT) criticality is a paradigmatic example of a phase transition beyond Landau's theory of symmetry breaking. The BKT universality class has been shown to underlie critical phenomena in numerous physical systems displaying superfluidity and superconductivity in thin films \cite{Bishop_1979, Bishop_1980}, Josephson junction arrays \cite{Resnick_1981, Newrock_2000}, ultracold gases \cite{Hadzibabic_2006,Murthy_2015}, polaritonic systems \cite{Nitsche_2014, Caputo_2018}, among others. 

The underlying physical mechanism of the BKT transition is a proliferation of vortices, acting as disorder operators in the statistical mechanics of two-dimensional $U(1)$ symmetric models \cite{Kosterlitz_1973,Jose_1977}. The transition is then driven by the subtle energy-entropy competition between the energy cost of unbinding opposite charge vortex pairs and the configurational entropy gain. Crucially, both follow logarithmic scaling with respect to the spatial separation between vortex anti-vortex pairs. The resulting phase diagram, when described in terms of a vortex Coulomb gas, features a low-temperature phase with confined, tightly bound vortex pairs and a high-temperature phase characterized by deconfined, free vortices.

\begin{figure}[h]
\captionsetup[subfigure]{labelformat=empty}
\subfloat[\label{subfig:bound_dipole}]{}
\subfloat[\label{subfig:free_charges}]{}
\subfloat[\label{subfig:sigma_phase_diagram}]{}
\subfloat[\label{subfig:rho_phase_diagram}]{}
\centering
\includegraphics[width=\columnwidth]{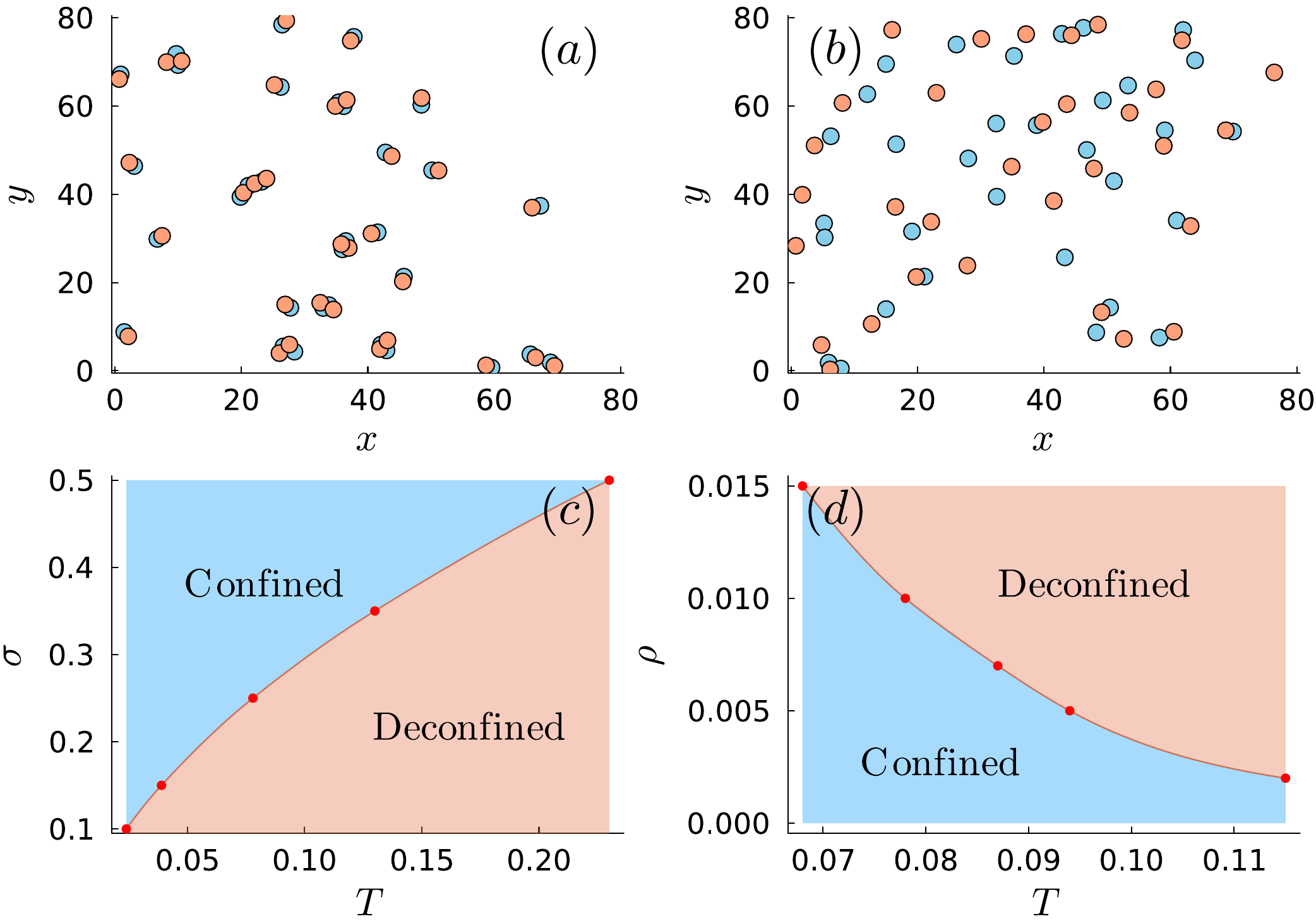}
\vspace{0mm}
\caption{\label{fig:phases} Monte Carlo snapshots of a super-Coulombic gas with confining power-law, $r^\sigma$, interaction, with $\sigma=0.25$ and particle density $\rho=0.01$ in its (a) confined and (b) deconfined phases separated by a BKT phase transition. The orange (blue) circles indicate positive (negative) charges. (c) The power law vs temperature phase diagram for fixed particle density $\rho=0.01$. (d) The particle density vs temperature phase diagram for fixed power law $\sigma=0.25$.  The red points indicate BKT critical points observed using our simulations; the phase boundary is constructed by interpolating between them.}
\end{figure}
 
 Recent experimental interest in systems with arbitrary long-range interactions \cite{Yan_2013, Bruzewicz_2019, Bendkowsky_2009, Dunning_2024, DeSalvo_2015} has posed questions about the fate of the BKT transition in the presence of a more general power-law interaction, $r^\sigma$ \cite{Giachetti_2021, Defenu_2023, Radzihovsky_2024, Giachetti_2023, Yao_2024, Xiao_2024}. In particular, two recent works \cite{Giachetti_2023,Radzihovsky_2024}, studying the long-wavelength physics of super-Coulombic gases, i.e., gases with longer-ranged than Coulomb interactions, proposed, by means of Renormalization Group (RG) and dielectric screening (with a length-scale dependent dielectric function) calculations, that generic super-Coulombic gases {\em universally} display a screening crossover to an effective Coulomb description. This emergence is also expected in higher dimensions \cite{Herbut_2003, Radzihovsky_2024}.

An intriguing corollary of the aforementioned Coulomb emergence is that two-dimensional super-Coulombic gases were predicted \cite{Radzihovsky_2024,Giachetti_2023} to display a BKT class confinement-deconfinement transition at finite temperatures. This prediction starkly contrasts with the expectation of confinement at all temperatures, based on a naive energy-entropy argument that neglects screening.

In this Letter, we employ numerically exact calculations to showcase the super-Coulombic to Coulomb crossover and the subsequent BKT confinement-deconfinement transition in a concrete microscopic model. We identify low (high) temperature confining (deconfining) phases, see \cref{subfig:bound_dipole,subfig:free_charges}. To probe the emergent Coulomb interaction at long distances, we introduce a numerical method for extracting the interaction between test charges and determining the associated RG flow. The effective Coulomb coupling displays the universal Nelson-Kosterlitz jump \cite{Nelson_1977} at the deconfinement transition point, as expected for the BKT universality. The emergent Coulomb coupling in the bound phase increases sharply at low temperatures, reflecting power-law screening \cite{Radzihovsky_2024}, unlike conventional Coulomb gases, where it saturates to its bare value. Lastly, by studying this problem for different power laws and densities, we establish the generality of Coulomb emergence across a wide range of microscopic parameters and interaction forms, as shown in \cref{subfig:sigma_phase_diagram,subfig:rho_phase_diagram}.

{\it Microscopic model}  -- We consider a two-dimensional super-Coulombic gas governed by the Hamiltonian:
\begin{equation}
\label{eq:hamiltonian}
    H= -K\sum_{i< j} n_{i}n_{j}\left|\frac{r_{i}-r_{j}}{a}\right|^{\sigma},
\end{equation}
where $\sigma > 0$ dictates the interaction power law, $K>0$ is the bare coupling constant. $n_{i}=\pm1$ are unit charges respecting the $\sum_{i}n_{i}=0$ charge neutrality constraint. $a$ is a microscopic cut-off scale that in this study is fixed by the diameter of hard-sphere charges. For the case of a Coulomb plasma (i.e. the limit $\sigma \to 0$), the interaction takes a logarithmic form, $\log(r/a)$.

{\it Energy entropy balance} -- It is insightful to revisit the conventional energy versus entropy balance argument, as is presented for the BKT transition in the context of a 2D Coulomb gas, but for the case of our model. Following \cref{eq:hamiltonian}, it can be inferred that separating a pair of bound super-Coulombic charges by a distance $R$ incurs an energy cost $\sim R^{\sigma}$. On the other hand, the additional degree of freedom in the form of their relative spatial distance results in an entropy gain that scales as $\sim \log(R)$. Since in the limit $R \to \infty$, $R^{\sigma} \gg \log(R)$, the free energy is always reduced by minimizing the energy at all temperatures. One may thus naively expect super-Coulombic plasmas to always exhibit a single confined phase, consisting of bound charges. This contrasts with Coulomb gases, where the logarithmic scaling of both energy and entropy results in a confinement-deconfinement BKT transition at finite temperatures.

The aforementioned argument is incorrect, as it neglects the predicted screening effects \cite{Radzihovsky_2024, Giachetti_2023} resulting in a super-Coulombic to Coulomb crossover, as will be demonstrated explicitly in this work.

{\it Effective Coulomb coupling --} The standard technique for studying the BKT transition in Coulomb gases involves tracking the effective Coulomb coupling as an order parameter. Typically, the renormalization of the bare Coulomb coupling is extracted in terms of the charge density-density correlation function \cite{SM, Orkoulas_1996}. However, for a super-Coulombic gas displaying Coulomb emergence, the bare Coulomb coupling or the charge densities of emergent Coulomb particles are hard to define precisely, rendering the standard approach inapplicable. To address this, we devise a method to directly extract the effective potential experienced by fixed test charges in the presence of a super-Coulombic plasma medium. 

To that end, we consider a pair of opposite static test charges positioned at $x_{1}, x_{2}$ in the presence of a neutral medium comprising $N$ mobile charges $\{y_{i}\}$. The total Hamiltonian can then be written as a sum of three contributions, $H(x_{1},x_{2})= H_{C}(x_{1},x_{2})+H_{M}(\{ y_{i}\})+H_{I}(\{x_{i}\},\{y_{i}\})$. The first two terms correspond to independent energy contributions of the test charges and the medium, respectively. The last term captures the response of the medium to the presence of test charges. With the above definition, we can identify the effective potential experienced by test charges by averaging over the dynamical medium degrees of freedom as follows,
\begin{equation}
\label{eq:effective_test_potential}
 e^{-\beta V_{\text{eff}}(|x_{1}-x_{2}|)} = \langle e^{-\beta(H_{C}+H_{I})}\rangle_{H_M},
\end{equation}
 where, $\beta=1/T$ is the inverse temperature. $\left\langle \mathcal{O}\right\rangle_{H_M}=\frac{1}{\mathcal{Z}}\int_{\{y_i\}} e^{-\beta H_M} \mathcal{O}$, where $\mathcal{Z}=\int_{\{y_i\}} e^{-\beta H_M}$, is the canonical partition function of the charges in the medium.

For the case of Coulomb emergence \cite{Radzihovsky_2024,Giachetti_2023}, we anticipate an asymptotic Coulomb-like form $e^{-\beta V_{\text{eff}}(r)}=e^{-\beta(\kappa n^{2}_{t}\log(r)+c)}$, where $c$ is a constant, $n_{t}$ is the magnitude of the test charges and $\kappa$ is the long wavelength effective Coulomb constant. For intermediate scales, we define a running Coulomb coupling $\kappa(r)$ by measuring $V_{\text{eff}}(r)$ at two reference points, $r$ and $r/\sqrt{2}$:
\begin{equation}
\label{eq:K_coulomb}
\kappa(r) \equiv \frac{1}{n^{2}_{t}} \left(\frac{V_{\text{eff}} (r) - V_{\text{eff}} \left(\frac{r}{\sqrt{2}} \right)}{\log(\sqrt{2})}\right)
\end{equation}
For a super-Coulombic gas displaying emergent Coulomb behavior, $\lim_{r \to \infty} \kappa(r)$ should either saturate to a constant $\kappa_{\infty}$ in the bound phase with $\kappa_{\infty}>4T$ or vanish in the unbound phase of a Coulomb plasma. The transition between the phases should occur at a temperature satisfying the universal BKT relation $T=\kappa_{\infty}/4$ \cite{Nelson_1977}. This behaviour contrasts with bare super-Coulombic potentials, where $\lim_{r \to \infty}\kappa(r)\sim r^\sigma$ diverges to infinity.

For our results, we perform Markov Chain Monte Carlo simulations \cite{Orkoulas_1996, Saito_1981} of a canonical ensemble of neutral 2D super-Coulombic plasmas with $N$ charges at fixed particle density $\rho$, implicitly fixing a simulation box size $L$. To avoid boundary effects, the potential is measured in the bulk of the box, with a ``boundary buffer" of length $L/3$. For our finite-sized system, we define the long-wavelength Coulomb coupling as $\kappa_{L}\equiv \kappa(L/3)$. All energies and distances are measured in units of the bare coupling $K$ and cut-off $a$, respectively. Additional details about the simulation and an explanation of how our observable, defined in \cref{eq:effective_test_potential}, is computed are presented in the Supplemental Materials.
\begin{figure}[t]
\captionsetup[subfigure]{labelformat=empty}
\subfloat[\label{subfig:potential}]{}
\subfloat[\label{subfig:potential_temperature}]{}
\centering
\includegraphics[width=\columnwidth]{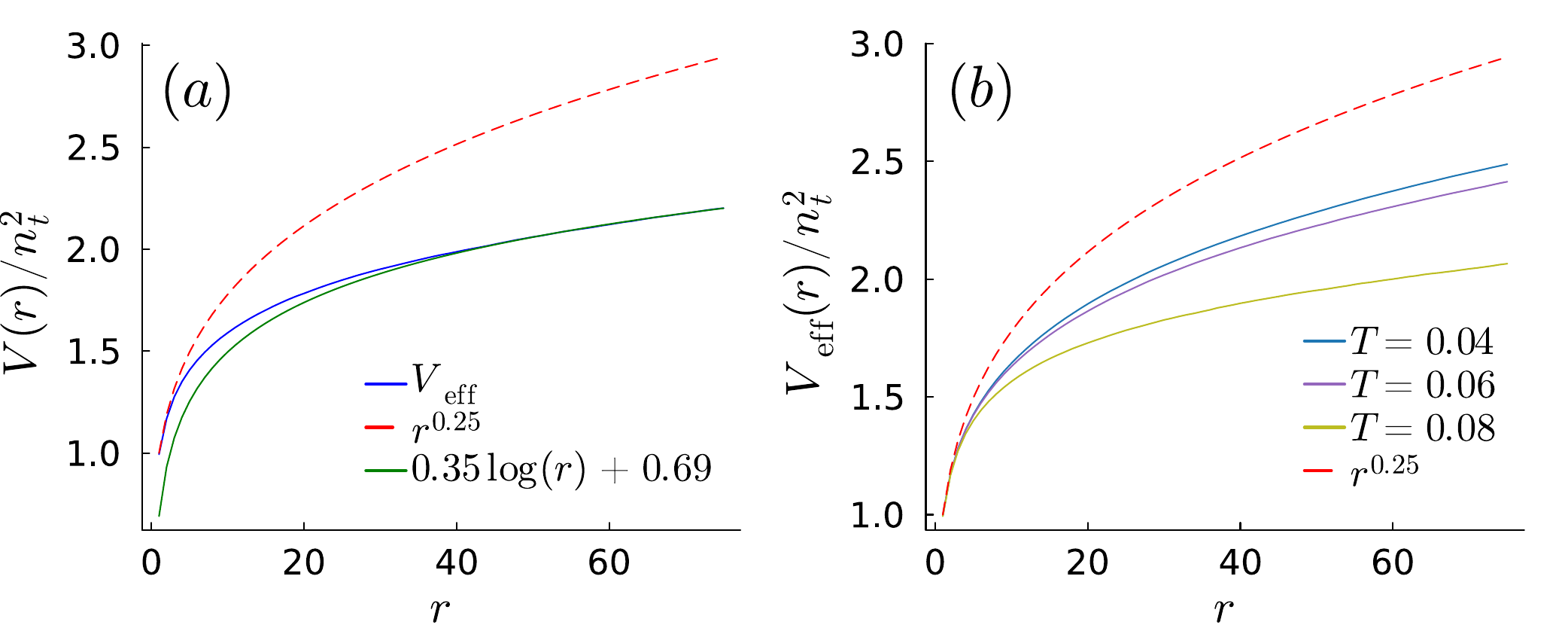}
\caption{\label{fig:potential} (a) The effective test charge potential for a $\sigma=0.25$, $\rho=0.01$, super-Coulombic gas at $T=0.075$, showcasing the crossover from its bare $r^\sigma$ potential to an emergent Coulomb-like logarithmic form. (b)  The effective potential for the same gas at different confining temperatures, contrasted against the bare interaction.}
\end{figure}

{\it Results -- }  In what follows, we present results for a $\sigma=0.25$, $\rho=0.01$ super-Coulombic gas unless specified otherwise. We begin by examining the functional form of $V_{\text{eff}}(r)$. By way of example, in \cref{subfig:potential}, we plot $V_{\text{eff}}(r)$ at a fixed temperature, contrasting it against the bare potential and an asymptotic Coulomb-like potential. We see that as $r$ increases, $V_{\text{eff}}(r)$ very quickly crosses over from its bare power-law form to a logarithmic Coulomb-like curve. We note that this is strong evidence for Coulomb emergence \cite{Radzihovsky_2024}. Moreover, in \cref{subfig:potential_temperature}, plotting $V_{\text{eff}}(r)$ for different temperatures, we see that the effective coupling decreases sharply with increasing temperature, reminiscent of Coulomb gas like behavior in the vicinity of a BKT confinement-deconfinement transition. 

 To observe signatures of the deconfinement transition, we plot $\kappa_{L}$ vs $T$ with varying system sizes in \cref{subfig:bkt_crossing}. We observe that $\kappa_{L}$ drops sharply as temperature increases. Looking at the scaling of $\kappa_{L}$ vs $T$ for different system sizes, we infer that thermodynamically, $\kappa_{\infty}$ takes a non-zero value at low temperatures and then sharply drops to zero beyond $T\gtrsim T_c=0.078(1)$. We attribute these two temperature regimes to the confined and deconfined phases, respectively. Monte Carlo snapshots of the super-Coulombic gas configurations as shown in \cref{fig:phases} further support the existence of bound and unbound charges in these temperature regimes, respectively. 

The transition between the two aforementioned phases is marked by a crossing point between the different finite system $\kappa_{L}$ vs $T$ curves, implying the existence of scale-invariant physics at this point. Remarkably, this point also lies on the $\kappa=4T$ universal BKT line, certifying its universality class. The thermodynamic extrapolation of crossing points between finite system curves determines $T_c$, as quoted above. In the supplementary \cite{SM}, we showcase this transition for different $\sigma$ and $\rho$ parameters, attesting that the $\kappa=4T$ relation holds across power-laws and densities, demonstrating the predicted universal occurrence of BKT transitions in generic super-Coulombic gases.

\begin{figure}
\captionsetup[subfigure]{labelformat=empty}
\subfloat[\label{subfig:bkt_crossing}]{}
\subfloat[\label{subfig:bkt_collapse}]{}
\centering
\includegraphics[width=\columnwidth]{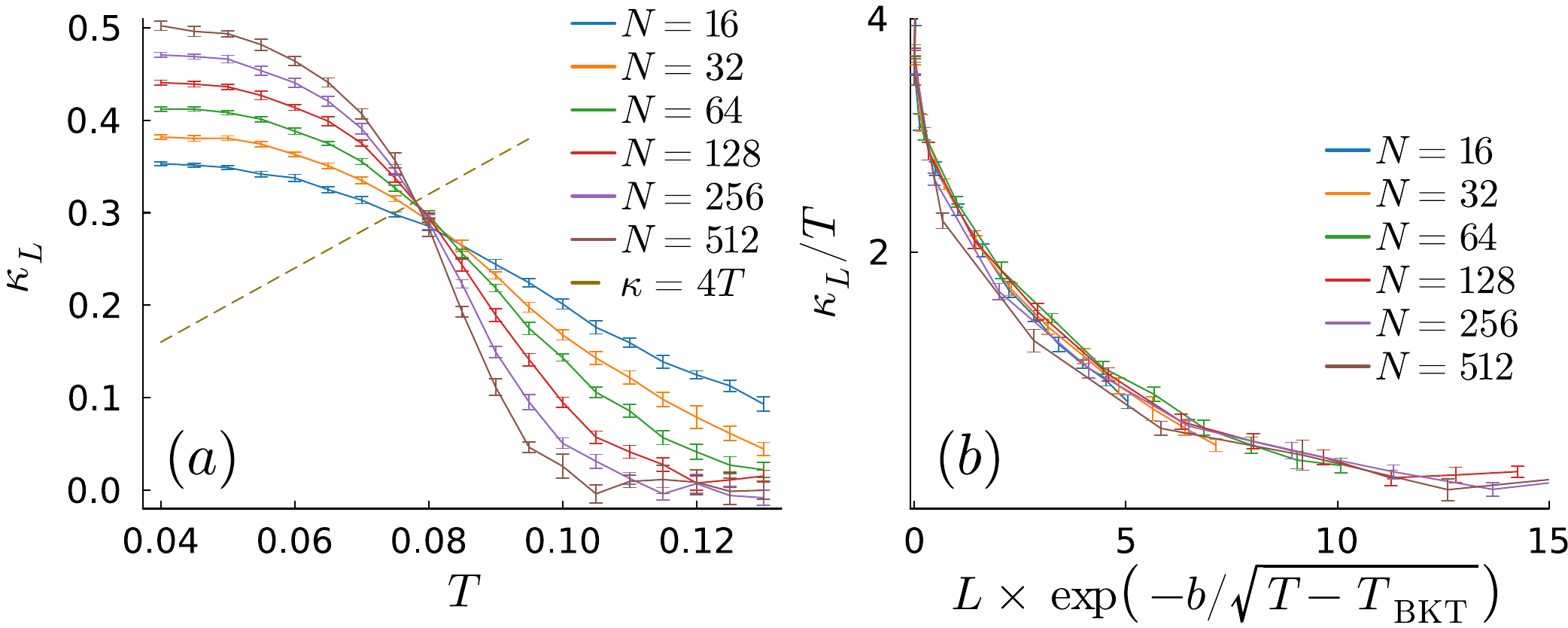}
\vspace{0mm}
\caption{\label{fig:bkt} (a) The Coulomb coupling vs temperature for a $\sigma=0.25$, $\rho=0.01$ super-Coulombic gas showcasing deconfinement. The dashed line represents the universal $\kappa=4T$ BKT line. (b) The universal BKT scaling function for $\kappa_{L}/T$ vs $L/\xi$ in the vicinity of $T>T_{\text{BKT}}$ for the same gas with activated scaling $\xi \sim \exp(b/\sqrt{T-T_{\text{BKT}}})$. Note that since $\rho$ is fixed, $L\sim \sqrt{N}$.}
\end{figure}

In \cref{subfig:bkt_collapse} we extract the universal scaling function associated with the dimensionless observable $\kappa_{L} / T$ near (above) the BKT point by plotting against the activated scaling variable $L\times\exp(-b/\sqrt{T-T_{\text{BKT}}})$. Here, $b=0.47(4)$ is a non-universal scaling parameter obtained via a curve fit. Indeed, we observe that curves belonging to different system sizes collapse onto a single universal curve consistent with the BKT scaling prediction.

\cref{subfig:bkt_crossing} shows that in the confined phase $\kappa_{L}$ gradually increases with the linear box size $L$, in contrast to the expected constant Coulomb coupling $\kappa$ in the thermodynamic limit. We attribute this to a finite-size RG flow associated with the super-Coulombic to Coulomb crossover, that saturates only at length scales greater than $L$. To support our claim, we study the scaling of $\kappa_{L}$ vs $1/L$ for different temperatures in the confined phase in \cref{subfig:coupling_scaling}. To quantify the thermodynamic convergence of $\kappa_{L}$ more accurately, we extract $\kappa_{\infty}$ by fitting our results to the ansatz $\kappa_{L}=\kappa_{\infty}-AL^{-\omega}$, where $A$ and $\omega$ are fitting parameters. The extrapolation as depicted in \cref{subfig:coupling_scaling}, shows that $\kappa_{L}$ indeed approaches a constant value $\kappa_{\infty}$ in the limit $L \to \infty$, suggesting Coulomb emergence. This convergence slows down at lower temperatures, as the screening length grows \cite{Radzihovsky_2024}, requiring even larger system sizes for $\kappa_{L}$ to saturate.

The temperature scaling of $\kappa_{\infty}$ in \cref{subfig:coupling_scaling} suggests a monotonic divergence of the effective Coulomb coupling with a decreasing temperature, deeper in the confining phase. This should be contrasted with the standard pure Coulomb case, which is expected to saturate to its bare value at low temperatures, as schematically depicted in \cref{subfig:coupling_schematic}. This behavior is consistent with the RG calculations in Ref. \onlinecite{Radzihovsky_2024}, which predict an exponential divergence of $\kappa_{\infty}$ at lower temperatures.

Since $\kappa_{L}$ flows for finite-sized systems, the length-scale at which the Coulomb regime fully settles is captured only when $\kappa_{L} \sim \kappa_{\infty}$. Nevertheless, it is fruitful to study the temperature dependence of the length-scale for which the crossover emerges, i.e., when $V_{\text{eff}}$ displays significant deviation from its bare form. In \cref{fig:potential}, we observe that for the specific choice of parameters, $V_{\text{eff}}$ departs from the microscopic super-Coulombic potential already at scales comparable to the microscopic cutoff. To resolve the evolution clearly, in Ref. \onlinecite{SM}, we study $V_{\text{eff}}$ at extremely low density, for which there is a separation of scales of the form $a\ll\xi_{c}\ll \rho^{-1/2}$. From this analysis, we observe that this length increases with decreasing temperature.

Alternatively, Ref. \onlinecite{Radzihovsky_2024} defines $\xi_{c}$ as the length-scale where the renormalized super-Coulombic contribution in $V_{\text{eff}}$ becomes comparable in magnitude to the Coulomb contribution generated under coarse-graining RG. Up to an $O(1)$ constant, this can be extracted using the relation $\xi_{c} \times O(1) = \kappa_{\infty}^{1/\sigma}$ \cite{Radzihovsky_2024}. In Ref. \onlinecite{SM}, we show that this length-scale diverges at low temperatures, similarly to $\kappa_{\infty}.$
\begin{figure}
\captionsetup[subfigure]{labelformat=empty}
\subfloat[\label{subfig:coupling_scaling}]{}
\subfloat[\label{subfig:coupling_schematic}]{}
\centering
\includegraphics[width=\columnwidth]{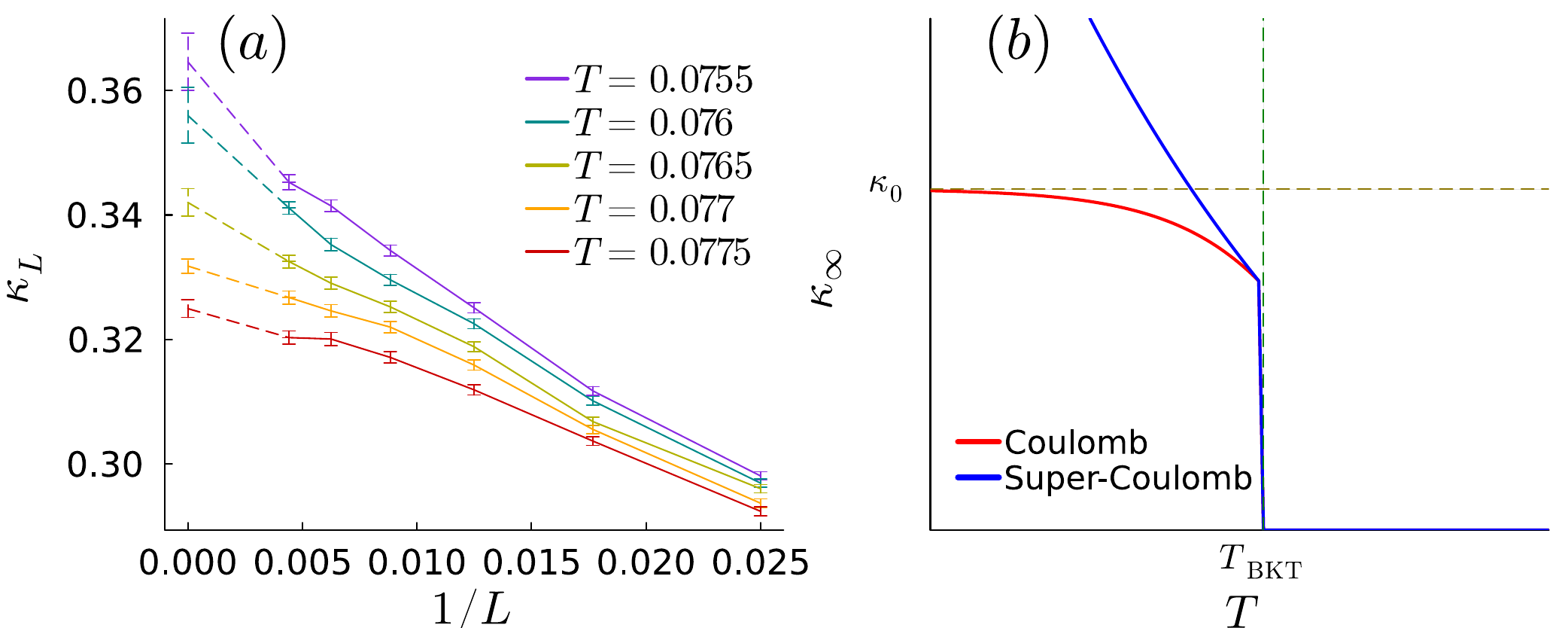}
\vspace{0mm}
\caption{\label{fig:coupling} (a) Finite size scaling of the effective Coulomb coupling for different temperatures in the confined phase. The dashed lines represent extrapolation to the thermodynamic limit, showcasing convergence to finite constants. (b) Schematic depiction of the temperature dependence of the thermodynamic Coulomb couplings for a microscopically Coulomb gas with bare coupling $\kappa_{0}$ and an arbitrary microscopically super-Coulombic gas.} 
\end{figure}

It is interesting to extend our study to driving the BKT transition by varying the power law $\sigma$ at a fixed temperature and density. In \cref{fig:sigma_rho}, fixing $T=0.078$ and $\rho=0.01$, we observe a deconfined phase with $\kappa=0$ for $\sigma \lesssim 0.25$, undergoing a transition to a confined phase with $\kappa \neq 0$ at a critical point $\sigma=0.25$, respecting the relation $\kappa=4T$. The phase diagram corroborates the intuitive expectation that gases with larger power laws bind charges more strongly, thus requiring higher temperatures to deconfine. Consequently, $T_{\text{BKT}}$ should increase with increasing $\sigma$. The BKT crossings shown in Ref. \onlinecite{SM} for different power laws exemplify this behavior.

Similarly, we investigate the phase diagram as a function of particle density $\rho$ for a fixed power law $\sigma$ and temperature $T$. It is known that in Coulomb gases, increasing the particle density decreases $T_{\text{BKT}}$ \cite{Orkoulas_1996}. In \cref{subfig:coupling_vs_density}, we show that super-Coulombic gases display similar behavior. Fixing $\sigma$ to 0.25 and $T$ to 0.078, we observe that as $\rho$ is increased, the gas undergoes a phase transition from a confined to a deconfined phase.  Moreover, in Ref. \onlinecite{SM} we show $\kappa_{L}$ vs $T$ BKT crossings with $\sigma=0.25$ for varying densities display an inverse relationship between $\rho$ and $T_{\text{BKT}}$.
Combining these results, we construct two-dimensional confinement-deconfinement phase diagrams in the $\sigma$ vs $T$ (\cref{subfig:sigma_phase_diagram}) and $\rho$ vs $T$ (\cref{subfig:rho_phase_diagram}) planes. 

\begin{figure}
\captionsetup[subfigure]{labelformat=empty}
\subfloat[\label{subfig:coupling_vs_sigma}]{}
\subfloat[\label{subfig:coupling_vs_density}]{}
\centering
\includegraphics[width=\columnwidth]{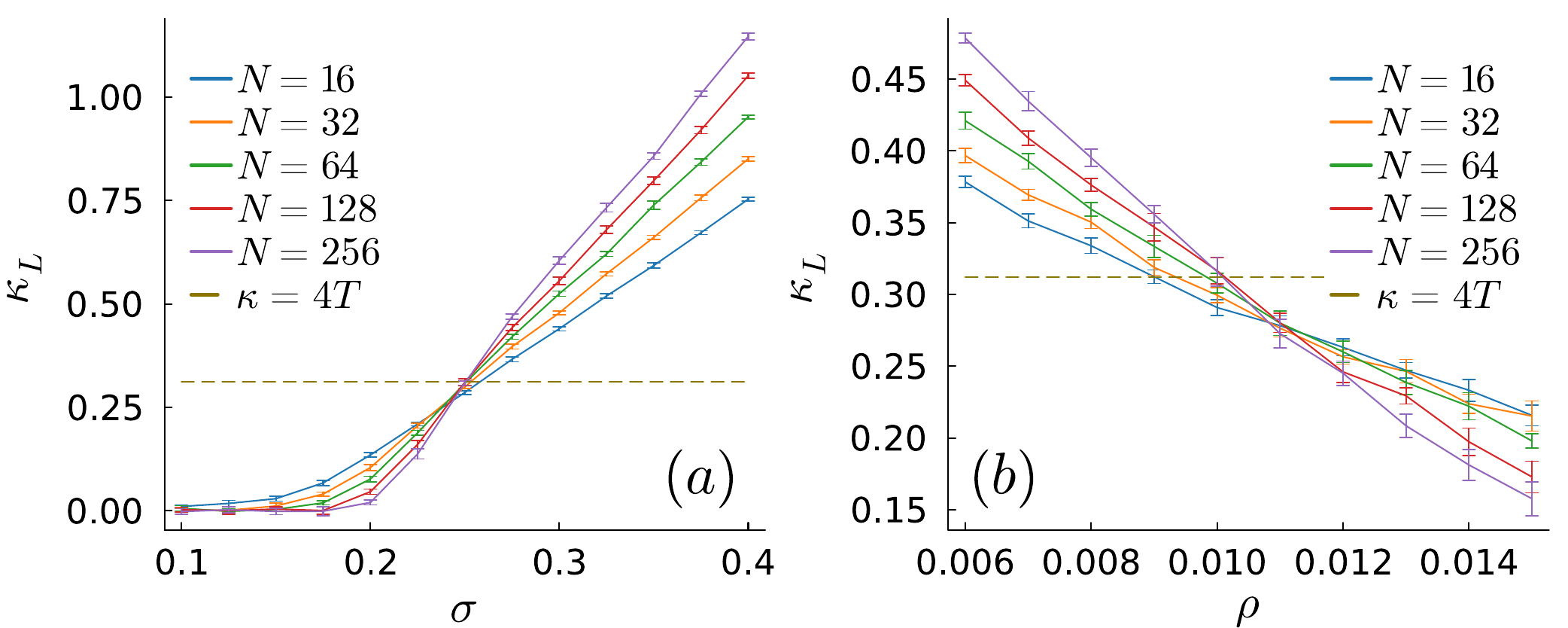}
\vspace{0mm}
\caption{\label{fig:sigma_rho} The Coulomb coupling for different system sizes showcasing the BKT transition by varying (a) the power law $\sigma$, and (b) the particle density $\rho$. Both plots are for a fixed temperature $T=0.078$. The dashed horizontal lines show the critical value $\kappa=4T_{\text{BKT}}\sim 4\times0.078$.}
\end{figure}

\emph{Summary and discussion --} Our results present direct and multifaceted numerical confirmation of novel dielectric screening in a two-dimensional super-Coulombic gas, showing the emergence of an effective Coulomb interaction beyond the screening length, corroborating recent RG predictions \cite{Radzihovsky_2024, Giachetti_2023}. This emergent behavior arises despite the underlying microscopic power-law interaction being longer-ranged than Coulomb, and is reflected in the asymptotic form of the effective potential beyond the crossover length scale $\xi_c$, and the saturation of the renormalized Coulomb coupling $\kappa_\infty$ at large distances.

A key physical consequence of this emergent Coulomb regime is the striking appearance of a finite-temperature confinement-deconfinement transition. We convincingly demonstrate that this transition lies in the BKT universality class, as evidenced by a universal jump in the effective coupling and finite-size scaling consistent with BKT behavior. Importantly, we find that this transition persists across a wide range of interaction power laws and particle densities, establishing its generality.

To characterize this crossover and transition, we introduced a test-charge-based method that extracts the effective interaction in the medium without relying on assumptions about microscopic charge densities or bare Coulomb couplings. This technique enables a robust and broadly applicable approach for probing emergent interactions in systems where standard correlation-function-based methods fail.

A notable observation is that the crossover into a true Coulomb description is long-tailed, saturating at distances much longer than length-scales where screening becomes significant. We defer a detailed study of this intermediate crossover regime to future work. Additionally, the predicted anomalous screening behavior in the deconfined phase, characterized by power-law rather than exponential (Debye-H{\"u}ckel) decay of interactions in Ref. \onlinecite{Radzihovsky_2024} remains an intriguing open direction for theoretical and numerical study. Finally, generalizing these ideas to higher dimensions, where Coulomb emergence is still expected \cite{Herbut_2003, Radzihovsky_2024}, presents a natural extension of this work.

\begin{acknowledgments}
\emph{Acknowledgments --} We thank Pradyumna Belgaonkar for useful discussions.  S.G. acknowledges support from the Israel Science Foundation (ISF) Grant no. 586/22 and the US–Israel Binational Science Foundation (BSF) Grant no.  2020264. Computational resources were provided by the Intel Labs Academic Compute Environment and the Fritz Haber Center for Molecular Dynamics, The Hebrew University of Jerusalem. L.R. thanks John Toner for discussions, feedback on the manuscript, and collaboration on Ref.~\onlinecite{Radzihovsky_2024}. L.R. acknowledges financial support by the Simons Investigator Award from the James Simons Foundation, and thanks Physics Department of The Hebrew University of Jerusalem for hospitality during a visit when this project was conceived. 
\end{acknowledgments}

\emph{Data Availability --} The data that supports our results can be found in Ref. \onlinecite{data}.
\bibliography{References} 
\fi
\ifSM
\widetext
\pagebreak
\begin{center}
\textbf{\large Supplemental Materials: Emergent Berezinskii-Kosterlitz-Thouless deconfinement in super-Coulombic plasmas}
\end{center}

\setcounter{equation}{0}
\setcounter{figure}{0}
\setcounter{table}{0}
\setcounter{page}{1}
\makeatletter
\renewcommand{\theequation}{S\arabic{equation}}
\renewcommand{\thefigure}{S\arabic{figure}}
\renewcommand{\bibnumfmt}[1]{[S#1]}
\renewcommand{\citenumfont}[1]{S#1}

\section{Berezinskii-Kosterlitz-Thouless Transition in a Coulomb Gas}

\begin{figure}[h]
\captionsetup[subfigure]{labelformat=empty}
\subfloat[\label{sm_subfig:coulomb_nkk_bkt}]{}
\subfloat[\label{sm_subfig:coulomb_test_bkt}]{}
\centering
\includegraphics[width=\columnwidth]{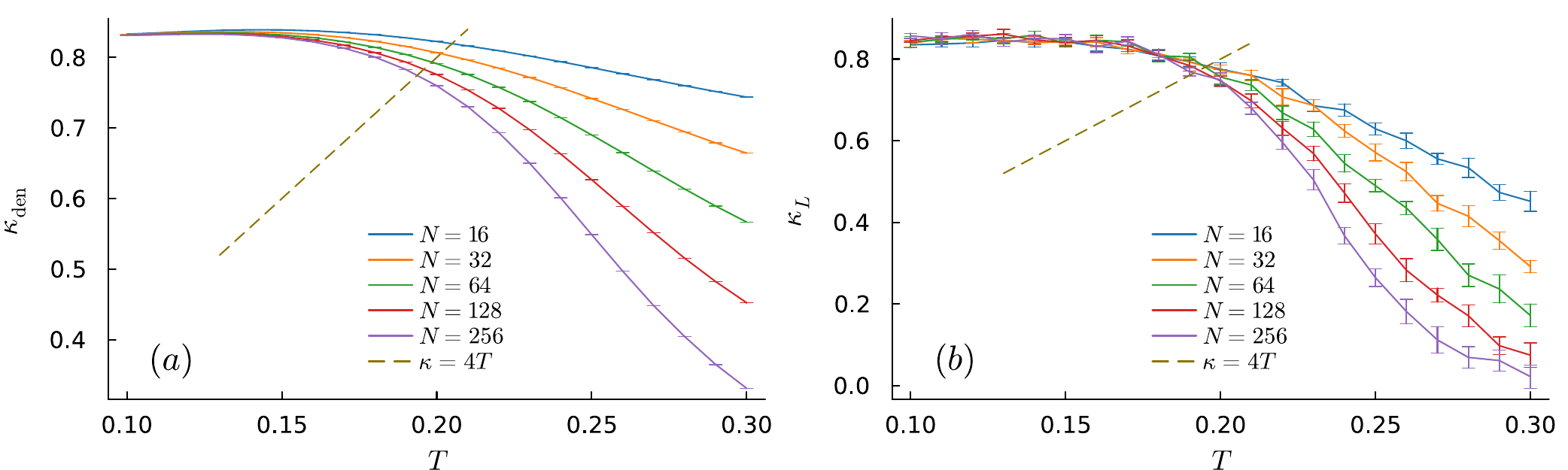}
\vspace{2mm}
\caption{\label{sm_fig:coulomb_bkt} The renormalized Coulomb coupling for a $\rho=0.0  1$ Coulomb gas in units of the bare coupling $K_{0}$ for different system sizes as a function of temperature extracted using (a) the relation in \cref{sm_eq:dielectric} and (b) using our test charge based approach as employed for super-Coulombic gases in the main text. To alleviate boundary effects due to the finite Monte Carlo box size, the low momentum limit for (a) is approached by taking $k=3\times2\pi/L$. The dashed lines in (a) and (b) mark the universal BKT constraint.} 
\end{figure}

As a validation step for our test charge-based approach, we study the BKT transition in a bare Coulomb gas. We compare this result with the standard method of calculating the effective Coulomb coupling using the charge density-density correlations given by the following relation arrived at using linear response theory \cite{Orkoulas_1996},
\begin{equation}
\label{sm_eq:dielectric} 
\kappa_{\text{den}}=\frac{K_{0}}{\epsilon}=\lim_{k \rightarrow 0} \left[ K_{0}-  \frac{2 \pi K^{2}_{0}}{Tk^2} \left\langle n(k)n(-k) \right\rangle \right].
\end{equation}
Where $K_{0}$ is the bare Coulomb coupling and $\epsilon$ is the dielectric constant. An equivalent relation for super-Coulombic gases showcasing emergent Coulomb behavior is unavailable because the bare Coulomb coupling $K_{0}$ and the density of emergent Coulomb charges are a priori not defined. 

In \cref{sm_fig:coulomb_bkt}, we compare the effective Coulomb coupling, extracted from both the density-density correlations (\cref{sm_subfig:coulomb_nkk_bkt}) and our test charge-based approach (\cref{sm_subfig:coulomb_test_bkt}). Indeed, we find that in both cases, the coupling flows to zero in the thermodynamic limit beyond a critical temperature $T \gtrsim T_{\text{BKT}}$, marking a deconfinement transition. We estimate $T_{\text{BKT}}\approx 0.19$ in both cases, using the crossing with the BKT line $\kappa=4T$.  However, we note that the precise finite system size results differ between the two methods. This can be attributed to the need to take the $k \to 0$ limit in \cref{sm_eq:dielectric}, which is only justified at infinite system sizes. Our test charge approach, on the other hand, shows faster convergence to the thermodynamic result.

\section{Signature of deconfinement in Super-Coulombic density-density correlations}
\begin{figure}[h]
\captionsetup[subfigure]{labelformat=empty}
\centering
\includegraphics[width=270px]{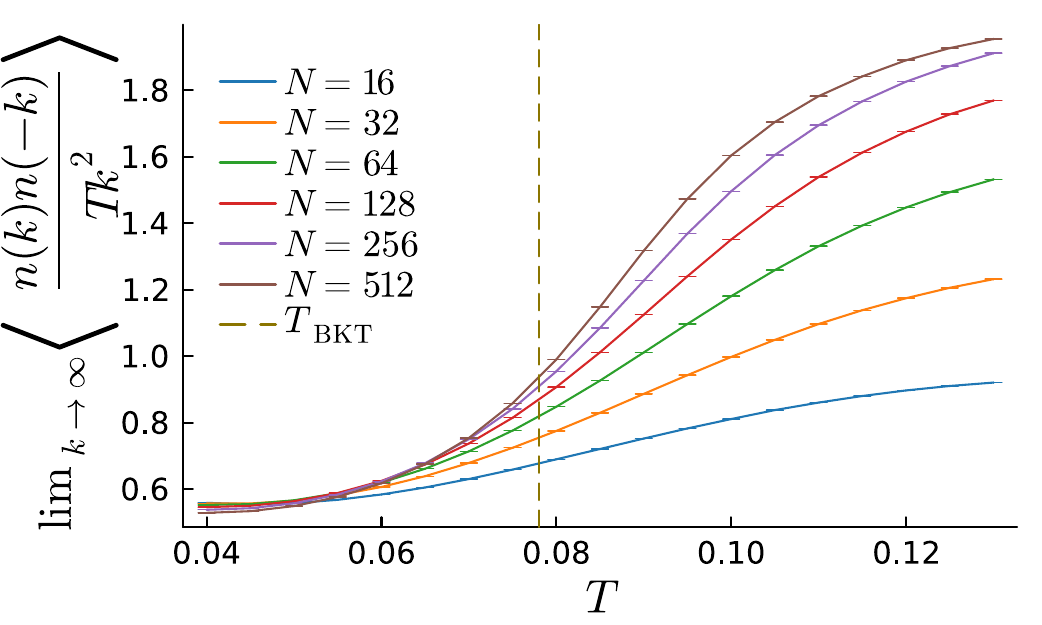}
\vspace{5mm}
\caption{\label{sm_fig:sc_density_density} The charge density-density correlation divided by $Tk^2$ in the long wavelength limit as a function of temperature for a $\sigma=0.25$, $\rho=0.01$ super-Coulombic gas. The dashed line marks the thermodynamic BKT temperature extracted from \cref{subfig:bkt_crossing}. To alleviate boundary effects due to the finite Monte Carlo box size, the low momentum limit is approached by taking $k=3\times2\pi/L$.} 
\end{figure}

BKT physics is directly encoded in the Coulomb charge density-density correlations.  For Coulomb gases, the low momenta correlations transition from scaling as $\lim_{k \to 0} \langle n(k)n(-k) \rangle \sim k^{2}$ for $T < T_{\text{BKT}}$ to $\lim_{k \to 0} \langle n(k)n(-k) \rangle \sim \frac{k^{2}}{k^2 + \xi^{-2}}$ for $T > T_{\text{BKT}}$. Here, $\xi$ marks the finite correlation length in the high-temperature deconfined phase. This is most evident in \cref{sm_eq:dielectric}, where the sudden drop in $\kappa$ at the BKT transition for Coulomb gases is captured by the change in functional form of the low-momentum charge density-density correlations. For super-Coulombic gases, however,  we do not have direct access to an equivalent emergent Coulomb charge density or the bare Coulomb constant.  Nevertheless, it is instructive to study the imprints of the confinement-deconfinement transition on the low-momentum charge density correlations of bare super-Coulombic particles. 

In \cref{sm_fig:sc_density_density}, we plot $\lim_{k\to 0}\langle n(k)n(-k)/(k^{2}T) \rangle$ for a super-Coulombic gas as a function of temperature for different system sizes. In the bound phase, for $T \leq T_{\text{BKT}}$, we see our observable takes a constant value for all system sizes. Since when approaching the limit $k \to 0$ with a fixed particle density $\rho$, the lowest discrete momentum $k \propto  L^{-1}$ has an implicit scaling with system size, the $N$ independence of our observable for $T<T_{\text{BKT}}$ implies that $\langle n(k)n(-k) \rangle \sim k^2$ in the bound phase. By contrast, the peeling off of the different curves for $T \gtrsim T_{\text{BKT}}$ suggests an abrupt change in the functional form of $\langle n(k)n(-k) \rangle$ across $T_{\text{BKT}}$, signaling charge deconfinement. 

\section{Super-Coulombic BKT for different power laws and densities}
\begin{figure}[h]
\captionsetup[subfigure]{labelformat=empty}
\centering
\includegraphics[width=450px]{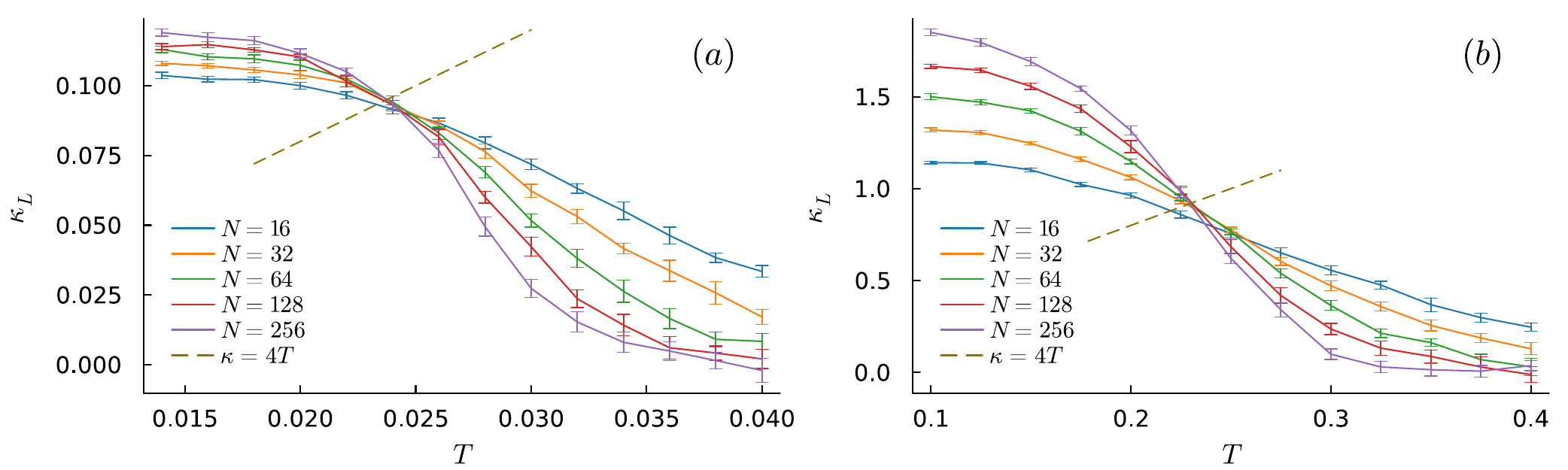}
\vspace{5mm}
\caption{\label{sm_fig:power_law_bkt} The emergent Coulomb coupling as a function of temperature for varying finite-sized systems with density $\rho=0.01$ for a super-Coulombic gas with power law (a) $\sigma=0.1$ and (b) $\sigma=0.5$. The dashed line indicates the universal BKT curve.}
\end{figure}

\begin{figure}[h]
\captionsetup[subfigure]{labelformat=empty}
\centering
\includegraphics[width=450px]{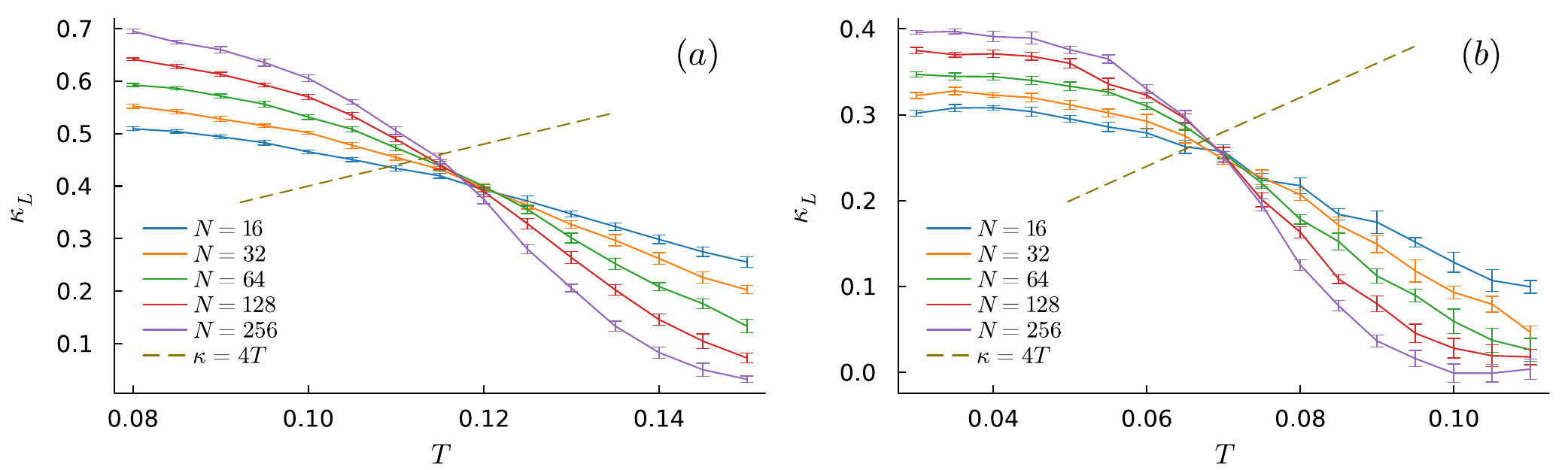}
\vspace{5mm}
\caption{\label{sm_fig:density_bkt} The emergent Coulomb coupling versus temperature for differently sized systems of $\sigma=0.25$ super-Coulombic gases with particle density (a) $\rho=0.002$ and (b) $\rho=0.015$. The dashed line indicates the universal BKT relation.} 
\end{figure}

To investigate the prevalence of BKT class confinement-deconfinement transitions in generic super-Coulombic gases, we look at the confinement-deconfinement crossings for different power laws in \cref{sm_fig:power_law_bkt} and particle densities in \cref{sm_fig:density_bkt}. For all the aforementioned cases, we observe that the scale-invariant crossing point marking the transition coincides with the universal BKT line. 

Also, we see in \cref{sm_fig:power_law_bkt} that the higher power law $\sigma=0.5$ shows a larger Coulomb coupling and consequently a higher BKT temperature $T_{\text{BKT}} \simeq 0.23$ as compared to the $\sigma=0.1$ case, where $T_{\text{BKT}}\simeq0.024$. 

Looking at \cref{sm_fig:density_bkt}, we see that, like in Coulomb gases, the lower density plot with $\rho=0.002$ transitions at a higher temperature $T_{\text{BKT}}\simeq0.115$ as opposed to $\rho=0.015$ where $T_{\text{BKT}} \simeq 0.068$.

\section{The Crossover Lengthscale}
\begin{figure}[h]
\captionsetup[subfigure]{labelformat=empty}
\subfloat[\label{sm_subfig:crossover_visual}]{}
\subfloat[\label{sm_subfig:crossover_xi}]{}
\centering
\includegraphics[width=\columnwidth]{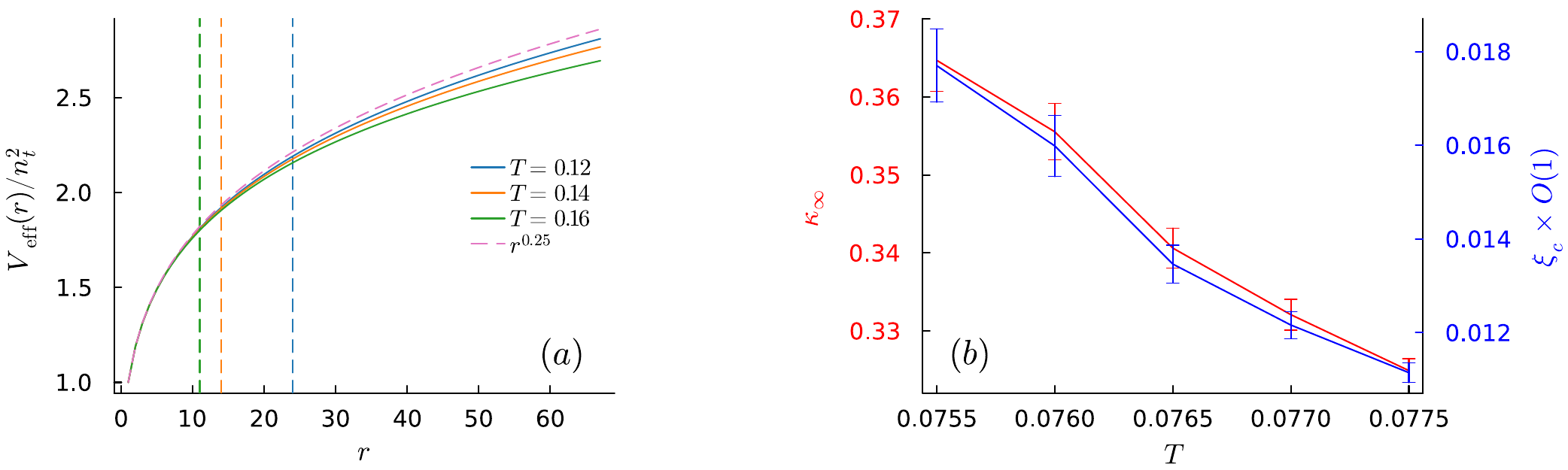}
\vspace{2mm}
\caption{\label{sm_fig:crossover} (a) The effective test potential in a $\sigma=0.25$ gas with a very low density $\rho=0.0002$ at three different temperatures. The vertical dashed lines show the points where the effective potentials deviate from the bare potential by more than one percent. (b) The thermodynamic coupling constant (left axis) $\kappa_{\infty}$ as extracted in \cref{subfig:coupling_scaling} and the associated crossover scale upto an $O(1)$ constant calculated using the relation $\xi_{c} \times O(1)=\kappa^{1/\sigma}_\infty$ (right axis).} 
\end{figure}
We now study the super-Coulombic to Coulomb crossover length scale. Due to the observed long-tailed nature of the crossover, the true Coulomb phase is only achieved when $\kappa_{L}$ saturates. Here, we focus on the length scale where screening effects become significant. To make a direct measurement of $\xi_{c}$, we consider a super-Coulombic gas at extremely low density where there is a clearer separation of scales between $\rho^{-1/2} \gg \xi_{c} \gg a$ in \cref{sm_subfig:crossover_visual}. We define $\xi_{c}$ as the distance where the effective potential deviates from it bare form by more than one percent, i.e., when $|V_{\text{eff}}(r)/n^{2}_{t}-r^\sigma|/r^{\sigma} \geq 0.01$. In \cref{sm_subfig:crossover_visual}, we use this metric to visually showcase $\xi_{c}$ at three different temperatures. It is evident that $\xi_{c}$
increases at lower temperatures.

Alternatively, using the relation $\xi_{c} \times O(1)=(\kappa_{\infty})^{1/\sigma}$ from Ref. \onlinecite{Radzihovsky_2024}, we extract $\xi_{c}$ upto an $O(1)$ constant in \cref{sm_subfig:crossover_xi} from the extrapolated $\kappa_{\infty}$ values in \cref{subfig:coupling_scaling}. We see that $\xi_{c}$ shows a divergence similar to $\kappa_{\infty}$ at lower temperatures. 

\section{Simulation Details}
\begin{figure}[h]
\captionsetup[subfigure]{labelformat=empty}
\subfloat[\label{sm_subfig:benchmark_energy}]{}
\subfloat[\label{sm_subfig:benchmark_nkkx_min}]{}
\centering
\includegraphics[width=450px]{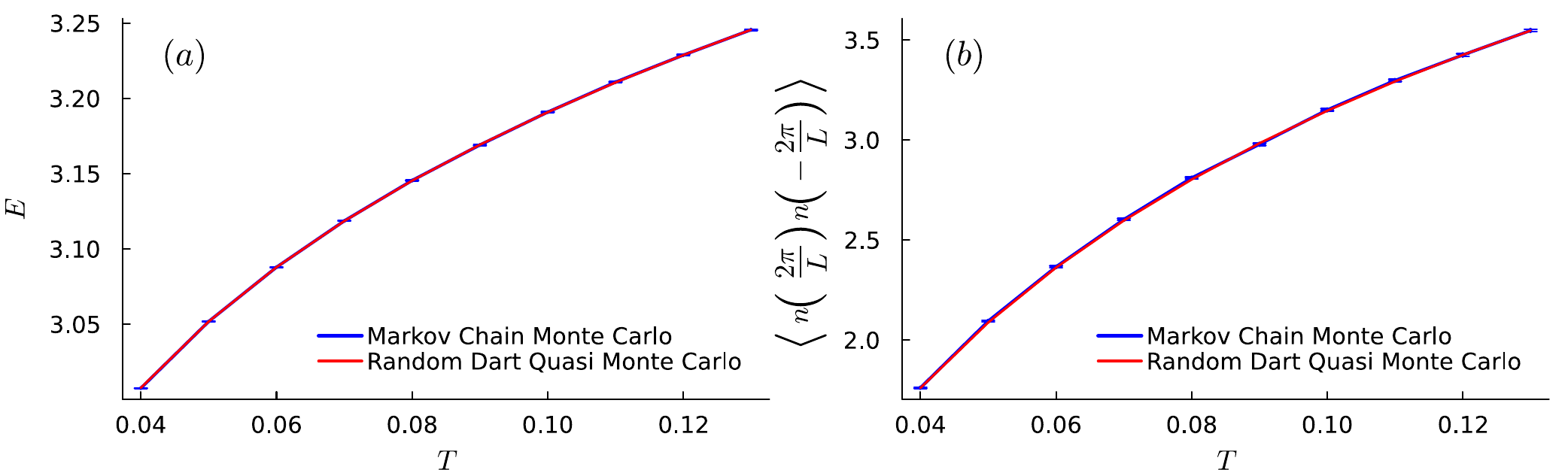}
\vspace{2mm}
\caption{\label{sm_fig:benchmark} Comparison of (a) the energy and (b) the charge density-density correlation as a function of temperature derived from our MCMC and a random dart quasi Monte Carlo procedure for a $\sigma=0.25$, $N=6$, $L=5.0$ super-Coulombic gas.} 
\end{figure}
Our Markov Chain Monte Carlo (MCMC) simulation samples from a canonical ensemble of unit hard sphere charges with interactions given by \cref{eq:hamiltonian}. Fixing the particle number $N$ and the density $\rho$ implicitly fixes the size of the simulation box. In our case, we work with open boundary conditions. 
The MCMC moves are designed to propose local changes to a given configuration. The transition probabilities are then determined via the Metropolis-Hastings algorithm. The moves come in two simple flavors: (1) \texttt{move particle }--  proposes to move a randomly chosen particle along a randomly chosen displacement vector. This move is dominant in the deconfined phase, where charges are independent. (2) \texttt{move dipole } -- randomly identifies a closely bound dipole pair and proposes moving it along a random direction. This move is important in the confined phase, where moving a single charge away from its dipole pair is energetically unfavorable.

To benchmark our simulation, we compare our MCMC results against a ``random dart" quasi-Monte Carlo calculation. Here, a configuration $c$ is drawn using quasi-random Sobol sequences \cite{Sobol_1967} and assigned a weight $e^{-\beta H(c)}$. The expectation values of observables are then computed as $\langle O \rangle = \frac{\sum_{c}O(c)e^{-\beta H(c)}}{\sum_{c}e^{-\beta H(c)}}$. A benchmark comparison between our MCMC and the random dart quasi-Monte Carlo methods is shown in \cref{sm_fig:benchmark}.

\section{Extracting the effective potential}

We now detail our numerical approach to computing $V_{\text{eff}}$ introduced in the main text. To that end, it is illuminating to rewrite \cref{eq:effective_test_potential} as:
\begin{equation}
\begin{aligned}
\label{sm_eq:effective_test_potential}
    e^{-\beta V_{\text{eff}}(x_{1},x_{2})} &= \langle e^{-\beta(H_{C}+H_{I})}\rangle_{H_M} \\&= \frac{\int_{\{y_{i}\}} e^{-\beta H}}{\int_{\{y_{i}\}}e^{-\beta H_{M}}} = \frac{\mathcal{Z}(x_{1},x_{2})}{\mathcal{Z}_{0}},
\end{aligned}
\end{equation}
where, $\mathcal{Z}(x_{1},x_{2})$ and $\mathcal{Z}_{0}$ are partition functions of the system in the presence and absence of test charges, respectively.

One way to compute the effective potential in \cref{sm_eq:effective_test_potential} is to consider fictitious test charges, i.e., test charges whose interaction energy doesn't feature in the detailed balance calculations of the simulation. In such a case, the observable $\langle e^{-\beta(H_{C}+H_{I})}\rangle$ gets averaged over the distribution $e^{-\beta H_{M}}/\mathcal{Z}_0{}$. Note that this allows us to compute the entire potential $V_{\text{eff}}(r) \forall r$ in a single measurement pass since a given medium realization is not tied to any particular configuration of test charges.

\begin{figure}[h]
\vspace{5mm}
\captionsetup[subfigure]{labelformat=empty}
\subfloat[\label{sm_subfig:comparing_methods_potential}]{}
\subfloat[\label{sm_subfig:comparing_methods_coupling}]{}
\centering
\includegraphics[width=\columnwidth]{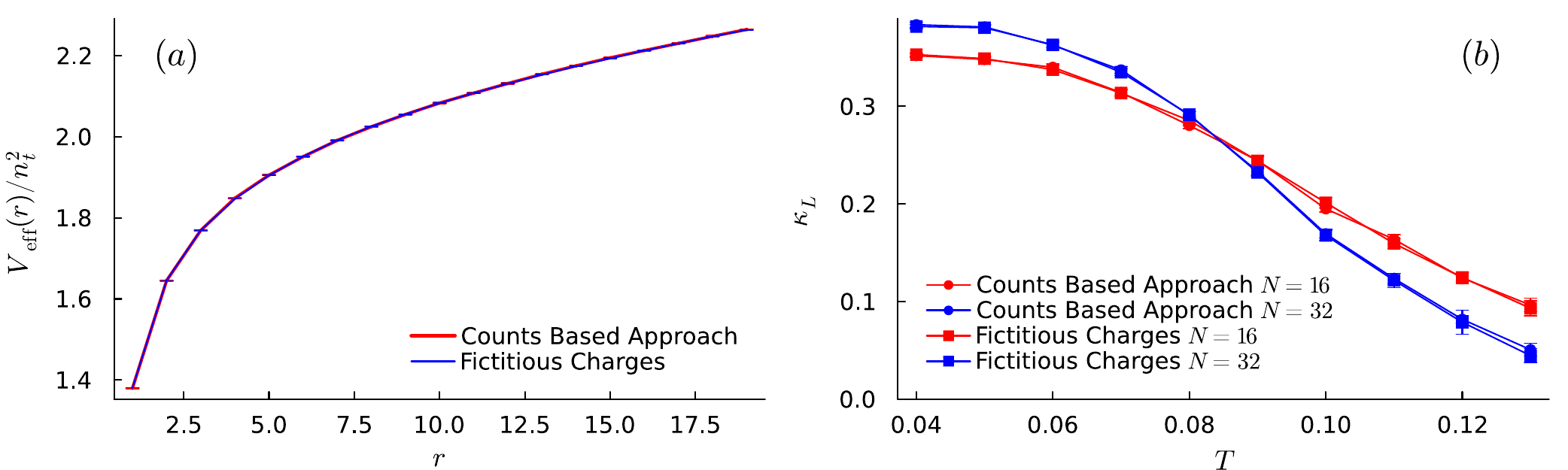}
\caption{\label{sm_fig:comparing_methods} (a) Comparison of the real space potential experienced by test charges extracted using the two outlined methods for $\sigma=0.25$, $\rho=0.01$, $N=32$ super-Coulombic gas at temperature $T=0.08$. (b) The emergent Coulomb coupling extracted using both methods for the same gas as a function of temperature.} 
\end{figure}

An alternate method is to sample over an ensemble of partition functions with test charges placed at varying distances. The combined partition function of this ensemble can be written as:
 \begin{equation}
 \label{sm_eq:total_test_part}
    \mathcal{G}=\sum_{\{r\}}\mathcal{Z}_{r},
 \end{equation}
where, $\mathcal{Z}_{r}$ is the partition function of a configuration with test charges fixed at two locations in the bulk separated by a distance $r$. $\mathcal{Z}_{0}$ is again the partition function of the medium in the absence of any test charge. Using \cref{sm_eq:effective_test_potential} and \cref{sm_eq:total_test_part}, $V_{\text{eff}}(r)$ can be computed from the sampled histogram of counts registered for the different $r$ sectors of the distribution $P(r)=\mathcal{Z}_{r}/\mathcal{G}$ as:
\begin{equation}
\label{sm_eq:count_test_potential}
e^{-\beta V_{\text{eff}}(r)}=\frac{\mathcal{Z}_{r}}{\mathcal{Z}_{0}}=\frac{\langle\delta_{r,r'}\rangle_{r'}}{\langle\delta_{0,r'}\rangle_{r'}}.
\end{equation}

This sampling approach is similar to the one employed for a different problem in Ref. \onlinecite{De_2024}. Since both our methods are equivalent, their comparison presented in \cref{sm_fig:comparing_methods} serves as a benchmark. Due to its better scalability, the fictitious charge approach is employed for the data in the main text.

\section{Test Charge Scaling}

\begin{figure}[h]
\captionsetup[subfigure]{labelformat=empty}
\subfloat[\label{sm_subfig:test_charge_potential_scaling}]{}
\subfloat[\label{sm_subfig:test_charge_coupling_scaling}]{}
\centering
\includegraphics[width=450px]{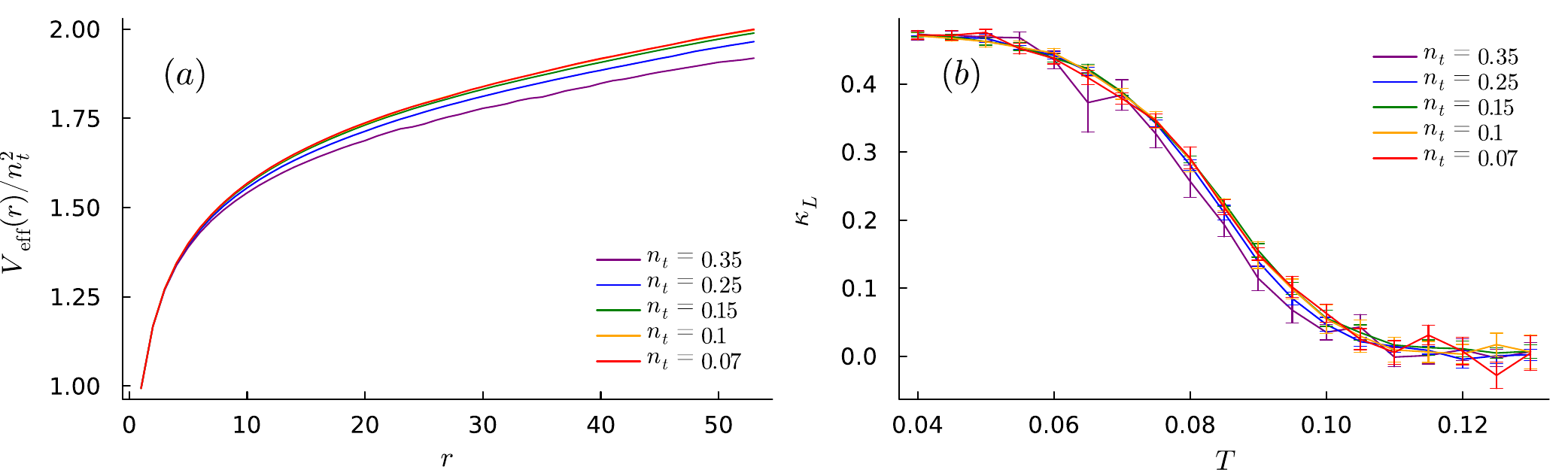}
\vspace{2mm}
\caption{\label{sm_fig:test_charge_scaling} (a) Comparison of the form of the effective potential for different magnitudes of test charges for a $\sigma=0.25$, $\rho=0.01$, $N=256$ super-Coulombic gas at temperature $T=0.08$. (b) The effective Coulomb coupling for the same gas as a function of temperature for different test charge magnitudes.} 
\end{figure}
Introducing test particles with a high magnitude of electrostatic charge can elicit non-linear responses from the medium, resulting in the ionization of dipoles in the medium. To ensure that we remain in the linear response regime, we track the convergence of the extracted potentials and Coulomb couplings in \cref{sm_subfig:test_charge_potential_scaling} and \cref{sm_subfig:test_charge_coupling_scaling} as we approach the small electrostatic charge ($n_{t} \to 0$) limit. While we see sizable nonlinearity for $n_{t}=0.35, 0.25$; $n_{t}=0.1$ is sufficiently small to be indistinguishable from a lower charge of $n_{t}=0.07$. Since smaller charges are susceptible to statistical noise, we proceed with $n_{t}=0.1$ for our analysis.

\section{Effect of hard-sphere defects}
\begin{figure}[h]
\captionsetup[subfigure]{labelformat=empty}
\subfloat[\label{sm_subfig:medium_density}]{}
\subfloat[\label{sm_subfig:medium_temp}]{}
\centering
\includegraphics[width=450px]{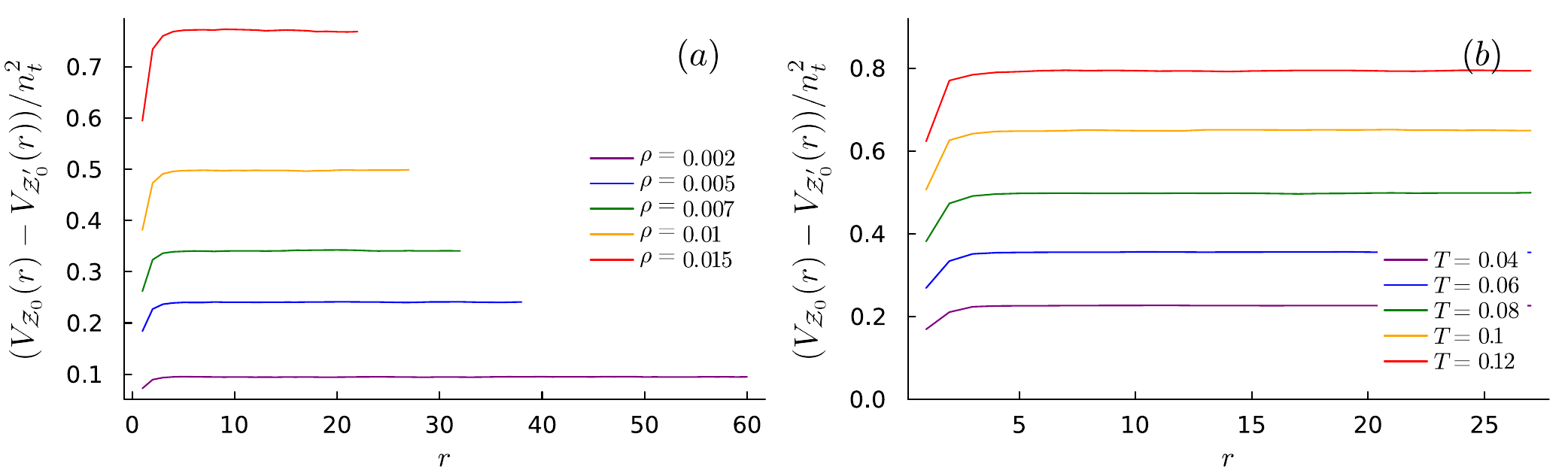}
\vspace{2mm}
\caption{\label{sm_fig:medium} Difference between the extracted potentials in $\sigma=0.25$ super-Coulombic gas mediums with and without defects at the location of the test charge for (a) different particle densities at a given temperature $T=0.08$, (b) different temperatures for given particle density $\rho=0.01$.  } 
\end{figure}
In this section, we highlight how at finite particle densities, the hard-sphere defects associated with the test charges affect the form of the extracted potential. While at large $r$ this results in a constant shift in the potential, at small distances this distortion is $r$ dependent.

To understand this, we consider a modified medium with two unit hard-sphere holes at the positions of the test charges (or, equivalently, a system where the electrostatic charge of the hard sphere test particles is set to zero). We call the partition function of this medium with hard sphere defects $\mathcal{Z}_{0}'(r)$, where $r$ is the distance between the defects. In the limit of high temperature and low density, the ratio of the partition functions can be shown to be:

\begin{equation}
\label{sm_eq:partition_ratio}
    \frac{\mathcal{Z}_{0}'(r)}{\mathcal{Z}_{0}} \simeq \left( \frac{\left(  \frac{L^2 - F(r) a^2}{\pi a^{2}} \right)^{N}}{\left(  \frac{L^2}{\pi a^{2}} \right)^{N}} \right) \simeq (1-F(r) \rho).
\end{equation}

where $\rho$ is the density and $F(r)$ captures the area of the excluded region for the medium charges due to the hard sphere defects. When the defects are sufficiently far away, this function takes a constant value, i.e., $F(r)=2 \pi$. On the other hand, when the defects are close enough that the excluded regions of both defects overlap, $F$ acquires an $r$ dependence. 

For a typical density of $\rho = 0.01$, $\mathcal{Z}_{0}'(r)/\mathcal{Z}_{0} \simeq 0.94$ at large $r$. Considering \cref{eq:effective_test_potential} and \cref{sm_eq:partition_ratio}, it is evident that depending on the choice of medium, $V_{\text{eff}}$ differs by $1/ \beta \log(1-F(r)\rho)$. While this difference is a constant at large $r$, at small $r$, it is $r$ dependent. 

Since the effective test charge potential must go to zero in the limit of $n_{t}=0$, the medium with defects truly isolates the electrostatic free energy of the test charges. On the other hand, the potential extracted from a defect-free medium also includes the free energy cost of introducing hard sphere defects into the medium. The difference between the two potentials is presented in \cref{sm_fig:medium}. 

When studying the form of the potential at short distances, like in \cref{fig:potential}, we simulate the medium with defects. For other cases, when computing the long-wavelength Coulomb coupling or charge density-density correlations, both choices of mediums yield identical results.

\section{Boundary Effects}
\begin{figure}[h]
\captionsetup[subfigure]{labelformat=empty}
\subfloat[\label{sm_subfig:finite_potentials}]{}
\subfloat[\label{sm_subfig:finite_pot_ratios}]{}
\centering
\includegraphics[width=450px]{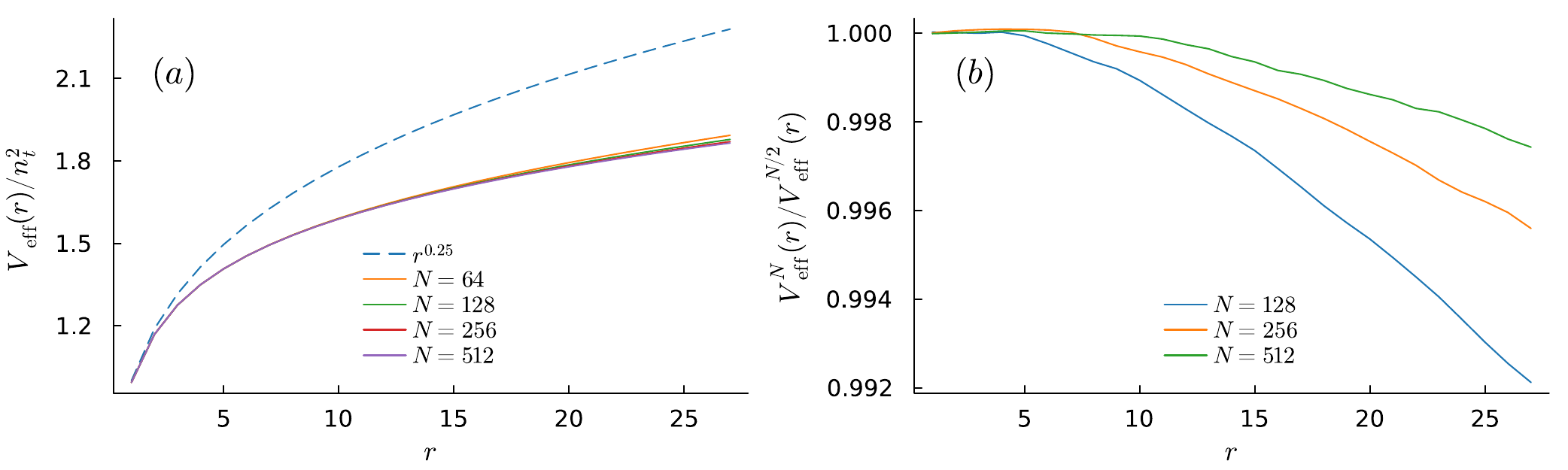}
\vspace{5mm}
\caption{\label{sm_fig:boundary_effect} (a) The form of the potential at $\rho=0.01$ and $T=0.075$ up to a fixed distance in the bulk for different system sizes, contrasted against the bare potential. (b) The ratio of potentials between successive system sizes.} 
\end{figure}

For systems with long-range interactions, as studied in this work, boundary effects can significantly affect emergent interactions in the bulk. To subvert this, as described in the main text, we choose an $L/3$ sized boundary buffer around the bulk region of study that approaches infinity in the limit of infinite box size. To show that this approach indeed systematically eliminates boundary effects, we study the potential up to a fixed distance $r_{\max}$ but for increasing system sizes in \cref{sm_fig:boundary_effect}. Since the particle density $\rho$ is constant, increasing $N$ increases the box size $L$, pushing the region of study deeper into the bulk. In \cref{sm_subfig:finite_potentials} we see that the potentials show minimal finite size effects, with successive system sizes converging to the same potential as $N$ approaches infinity. In contrast, the bare potential shows a drastically different functional form. In \cref{sm_subfig:finite_pot_ratios} we plot the ratio between the potentials obtained from systems of size $N$ and $N/2$. We see that for all $r$, this ratio systematically goes to $1$ as $N$ approaches the thermodynamic limit.

\ifmainText
\else
\bibliography{References} 
\fi
\fi
\end{document}